# A Semi-Classical View on Epsilon-Near-Zero Resonant Tunneling Modes in Metal/Insulator/Metal Nanocavities

*Vincenzo Caligiuri\*[†], Milan Palei[†§], Giulia Biffi[†§], Sergey Artyukhin[†], and Roman Krahne\*[†]*

[†] Istituto Italiano di Tecnologia, Via Morego 30, 16163 Genova, Italy

[§]Dipartimento di Chimica e Chimica Industriale, Università degli Studi di Genova, Via Dodecaneso, 31, 16146 Genova, Italy

AUTHOR EMAIL ADDRESS roman.krahne@iit.it

**RECEIVED DATE (to be automatically inserted after your manuscript is accepted if required according to the journal that you are submitting your paper to)**

CORRESPONDING AUTHORS:  Roman.krahne@iit.it; Vincenzo.caligiuri@iit.it

ABSTRACT

Metal/Insulator/Metal nanocavities (MIMs) are highly versatile systems for nanometric light confinement and waveguiding, and their optical properties are mostly interpreted in terms of surface plasmon polaritons. Although classic electromagnetic theory accurately describes their behavior, it often lacks physical insight, letting some fundamental aspects of light interaction with these structures unexplored. In this work, we elaborate a quantum mechanical description of the MIM cavity as a double barrier quantum well. We identify the square of the imaginary part $\kappa$ of the refractive index $n$ of the metal as the optical potential, and find that MIM cavity resonances are suppressed if the ratio $n/\kappa$ exceeds a certain limit, which shows that low n and high $\kappa$ are desired for strong and sharp cavity resonances. Interestingly, the spectral regions of cavity mode suppression correspond to the interband transitions of the metals, where the optical processes are intrinsically non-Hermitian. The quantum treatment allows to describe the tunnel effect for photons, and reveals that the MIM cavity resonances can be excited by resonant tunneling via illumination through the metal, without the need of momentum matching techniques such as prisms or grating couplers. By combining this analysis with spectroscopic ellipsometry on experimental MIM structures, and by developing a simple harmonic oscillator model of the MIM for the calculation of its effective permittivity, we show that the cavity eigenmodes coincide with low-loss zeros of the effective permittivity. Therefore, the MIM resonances correspond to *epsilon-near-zero* (ENZ) eigenmodes that can be excited via *resonant tunneling*. Our approach provides a toolbox for the engineering of ENZ resonances throughout the entire visible range, which we demonstrate experimentally and theoretically. In particular, we apply our quantum mechanical approach to asymmetric MIM superabsorbers, and use it for configuring broadly tunable refractive index sensors. Our work elucidates the role of MIM cavities as photonic analogues to tunnel diodes, and opens new perspectives for metamaterials with designed ENZ response.

Near zero dielectric permittivity is an unusual property often encountered in the framework of photonics.[1–6] The associated propagation regime, also known as *Epsilon-Near-Zero* (ENZ), is the technological framework in which several exciting phenomena have been observed, such as electric levitation,[7] energy squeezing,[8,9] sophisticated phase engineering of the electromagnetic radiation,[2] coherent perfect absorption[10] and Purcell effect enhancement.[11] ENZ conditions occur naturally in some noble metals, for example in Ag at 327 nm, and in polar crystals at optical phonon frequencies[5] and have been be referred to as Ferrell-Berreman modes.[12–15] However, designing materials with a customizable ENZ response is highly challenging. One successful approach with the ENZ wavelength in the infrared is constituted by conductive oxides like indium-tin-oxide (ITO) and aluminum doped zinc oxide (AZO). These materials possess intrinsic ENZ characteristics at given wavelengths in the IR, which are tunable via the concentration of their dopants, and they manifest strong non-linear effects.[11,16–20] A broad tunability of the ENZ response has not been achieved so far, and is especially challenging in the visible range. A promising approach consists in artificially engineering structures whose optical response can be easily manipulated by acting on the optical and geometrical properties of their fundamental components. This is the case for *Metal-Insulator-Metal* (MIM) multilayers.[21–29] MIM nanoresonators have been widely investigated in the past, since their straightforward fabrication makes them an ideal platform for the study of sophisticated plasmonic systems, such as negative index materials,[30,31] planar plasmonic waveguides,[32,33] color filters and superabsorbers.[34] Recently, it has been experimentally demonstrated that MIM nanocavities manifest ENZ response at their resonant modes,[35] which has been exploited to enhance the photophysical properties of fluorophores interacting with the MIM in a weak coupling regime.[35,36] Despite the experimental observations, the physical nature of the *epsilon-near-zero* response in MIMs remains still obscure.

In this work, we analyze the properties of MIM cavities with two novel and complementary approaches. First, we show that the MIM can be seen as a quantum mechanical potential well, with even and odd eigenmodes that correspond to tunneling maxima. Then we describe the MIM as a classical harmonic oscillator and derive a simple and useful equation for its effective dielectric permittivity, relating the eigenmodes and tunneling maxima with the ENZ frequencies. Rationalizing the MIM cavity resonances as ENZ and resonant tunneling modes straightforwardly elucidates that they can be excited without the need of a grating coupler or prism for momentum matching, as it would be required for coupling to surface plasmon polaritons.[37,38] The homogenization of the MIM's dielectric permittivity allows us to treat it as an effective potential barrier, reducing the complex problem of the propagation of the photon through a real double potential barrier to tunneling through a single barrier whose height is given by the square of the effective refractive index. The vanishing wavevector at the ENZ wavelength reduces the wavefunction of the photon within the barrier to a constant, enabling *resonant tunneling* of photons, in analogy to an electron impinging on a potential barrier with energy equal to its height. The models are applied to physically fabricated samples with silver (Ag) as metal and alumina (Al$_2$O$_3$) as dielectric, and compared with classical electromagnetic theory. We obtain a perfect match between all approaches and experiments that corroborates the quantum mechanical and harmonic oscillator interpretations. We put the classical-quantum analogy for metal/dielectric systems on solid ground by identifying the square of the imaginary part of the refractive index of the metal as the optical potential, and by discussing the necessary conditions to obtain a Hermitian system that is essential for quantum mechanical treatment.

Results and Discussion

The wave nature of the propagation of light in classical optics, and of electrons in quantum mechanics can lead to several analogies between these two fields that can provide additional insights. A classic textbook example is the propagation of light in a medium with refractive index $n$ (typically glass) under total internal reflection conditions across a thin air gap that leads to frustrated internal reflection.[39] This behavior is to some extent analogous to the tunneling effect in quantum mechanics, and we can derive a similarity between the time-independent Schrödinger equation, expressed as

$$\frac{d^2}{dx^2}\Psi(x) - \frac{2m}{\hbar^2}(U(x) - E)\Psi(x) = 0; \qquad (1)$$

and the Helmholtz equation

$$\frac{d^2}{dx^2}\Psi(x) + \left[k_0\left(n_j - i\kappa_j\right)\right]^2\Psi(x) = 0; \qquad (2)$$

by choosing an optical potential whose associated wavevector is a function of the complex refractive index of the $j^{th}$ material in which the propagation takes place (see Supporting Information (SI) section 1).[40,41] The wavevector $\boldsymbol{k}$ of the photons travelling in the material perpendicular to the interface can be expressed as $k = k_0\left(n_j(x) - i\kappa_j(x)\right)$, where $k_0 = \frac{2\pi}{\lambda}$ is the wave vector in the vacuum, $\tilde{n}_j = n_j(x) - i\kappa_j(x)$ is the complex refractive index of the $j^{th}$ material composed by a real part $n_j(x)$ and an imaginary part $\kappa_j(x)$. In eq (1) $m$ is the electron mass, $U$ the potential, and $E$ the energy.

Such a similarity has been fruitfully used, for example, in describing the properties of a particular class of non-Hermitian systems constituted by the so-called *parity-time symmetric* potentials.[42–47] The study of these systems unveiled the role of the square of the refractive index as the "optical potential" acting on photons. The further expansion of the quantum analogy to metal-insulator systems encounters the problem of hermiticity of the Hamiltonian, that is required to obtain physical solutions, but which is not necessarily satisfied for the given material system.

In particular, the Hamiltonian

$$H = \frac{d^2}{dx^2} + \left[ k_0 \left( n_j - i\kappa_j \right) \right]^2 \tag{3}$$

associated to eq 2 is non-Hermitian. However, if either the real or the imaginary part of the refractive index are negligible, then the second term in eq 3 becomes a real number, and hermiticity is fulfilled. Concerning metals and dielectrics, the imaginary part of the refractive index of most of the dielectrics is negligible, while for plasmonic metals the imaginary part dominates.[a] In dielectrics, the wavefunction is a propagating wave $\Psi_d(x) = \mathrm{Ae}^{-\mathrm{i}k_0 n_j x}$, while in metals it is essentially an exponentially decaying wave $\Psi_m(x) = \mathrm{Ae}^{-k_0 \kappa_j x}$. Consequently, the wave vector is $k_d = k_0 n_d$ for a dielectric and $k_m = \mathrm{i}k_0 \kappa_m$ for a metal. This reciprocal behavior ensures hermiticity in eq (2), and justifies the quantum mechanical analogy for many MIM systems. However, not all the metals have a refractive index that ensures hermiticity, as we demonstrate in Figure 1. For metals, where the ratio between the real and imaginary parts of the refractive index exceeds a certain value, hermiticity is strongly violated and the corresponding eigenvalues acquire significant complex parts. From Scattering Matrix Method (SMM)[49–51] modeling we find that the critical ratio is at 0.2 (+/-0.05); above this value, the resonant cavity modes cannot be sustained. This restricts the spectral range in which resonances can be engineered, as shown in Figure 1a for different metals. For Ag (black curve), the ratio $n/\kappa$ is always smaller than 0.2 in the visible range, which makes this material an ideal candidate for the study of ENZ resonances. In the case of Au (green curve), the ratio $n/\kappa$ is very large in the inter-band transition range, while for Al (red curve), the non-Hermitian band lies within its well-known plasmon absorbance band, at 675 - 1000 nm. Interestingly, Mg (blue curve) is Hermitian throughout the visible range. Figures 1b-d, show the SMM simulated absorbance for MIMs made of Au (b), Al (c), and Mg (c) for different dielectric layer thickness.[49-51] As predicted, resonances are quenched in the portions of the electromagnetic spectrum where the ratio $n/\kappa$ is larger than 0.2. Interestingly, the non-Hermitian range can be related to the interband transition energies of the metal, which explains the different spectral Hermitian/non-Hermitian regions for the various metals in Figure 1. In the interband transition range, a photon impinging on the metal is annihilated, exciting electron-holes pairs, and this intrinsically non-Hermitian phenomenon is not contemplated in the quantum mechanical picture and the Schrödinger equation. For example, in the case of Au, the cavity resonances are strongly quenched for wavelengths smaller that 500 nm, and it is impossible to distinguish well-defined Lorentzian absorbance peaks. The same happens for Al at wavelengths exceeding 700 nm. The case of the MIM with Al as metal with 230 nm dielectric layer thickness is particularly interesting (Figure 1c, magenta curve). Here, the odd mode falls within the Hermitian band and is well distinguishable as a sharp Lorentzian absorbance peak, while the corresponding even mode that lies in the non-Hermitian band is not observed. In the quantum mechanical picture, the $n/\kappa$ ratio gives a measure of how well the Schrödinger equation approximates the Helmholtz equation in the metal, by considering how well the wavefunction can be described by an evanescent wave.

In the following we will only consider MIM structures with Ag as a metal and $Al_2O_3$ as a dielectric.

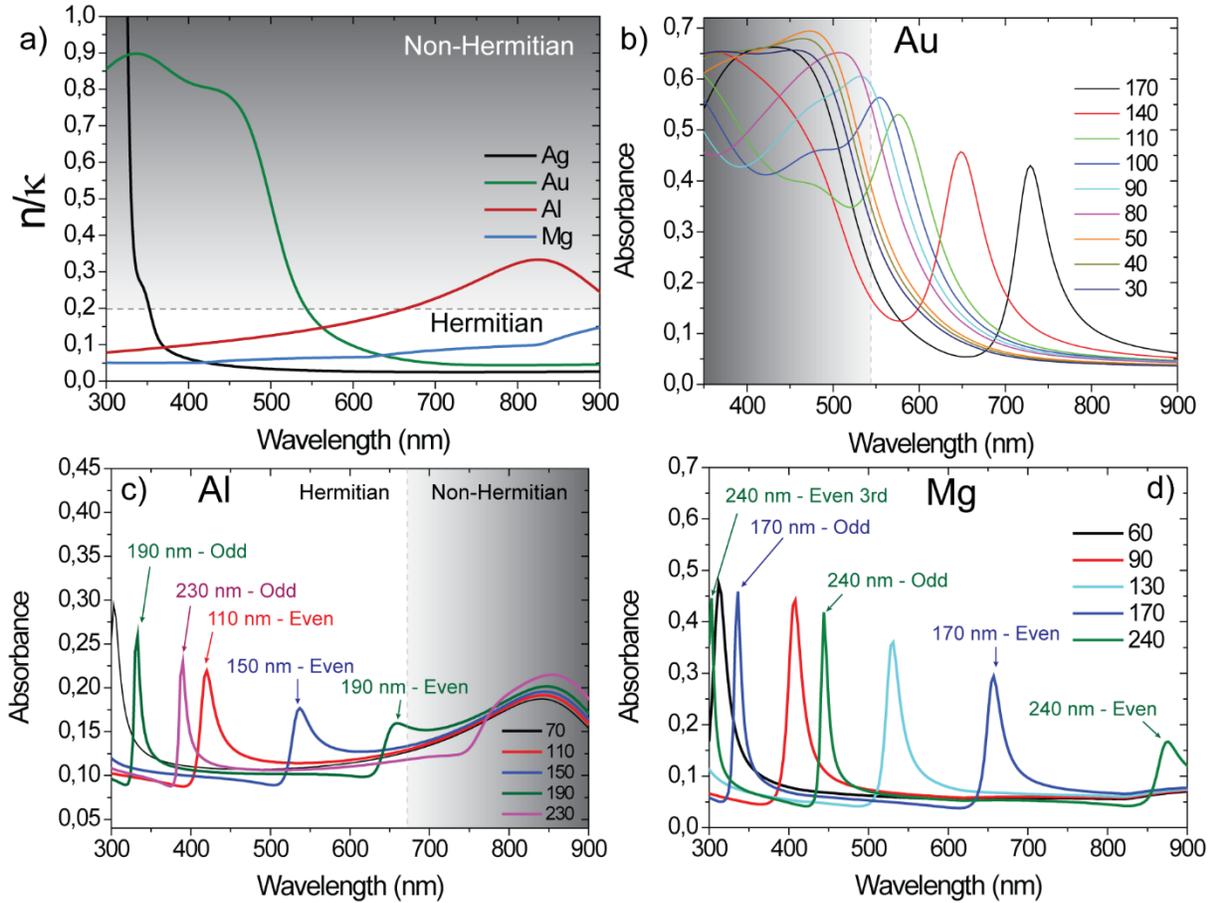

**Figure 1**. (a) Ratio n/$\kappa$ for the most common plasmonic metals defining the "hermiticity limit", in which ENZ resonances can be sustained in MIMs. SMM calculated absorbance (obtained as 1-transmittance-reflectance) for MIMs using (b) Au, (c) Al, and (d) Mg as metals, for different thickness of the Al$_2$O$_3$ dielectric layer.

MIM nanoresonators are the nanometric equivalent of the dielectric *Fabry-Perot cavities*.[32,34,52,53] As such, they can be treated as optical cavities with discrete resonant modes, albeit with a very short optical path. Within the quantum mechanical analogy, we can study a MIM system as a square potential well, where the metals constitute the barriers and the dielectric the well. A special case is a MIM cavity whose metallic layers that are much thicker than the skin depth, which can be treated as a finite square potential well, as depicted in the sketch of Figure 2a.

The height of the barrier is given by $\kappa_m^2$. The wavefunction of the cavity mode exponentially decays to zero in the thick metals, while a standing wave is formed inside the cavity.

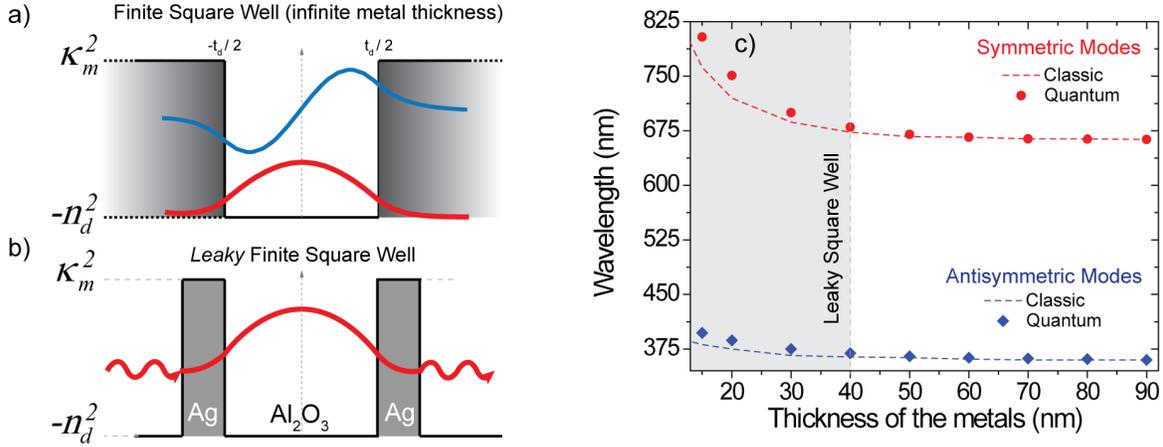

**Figure 2.** (a) Sketch of the square quantum well that is the quantum analogue of a MIM system with metallic layers much thicker than the skin depth. The corresponding experimental system consists of Ag layers of 100 nm thickness, and Al₂O₃ as a dielectric. Even and odd eigenmodes are shown in red and blue, respectively. (b) For thin metallic layers, a double barrier system with a finite tunneling probability through the barriers has to be considered. This "leaky" system confines "quasi-bound" eigenmodes. (c) Resonance wavelength of the even and odd eigenmodes, calculated via eqs 6 and 7, versus the metallic layer thickness for a symmetric MIM. The dielectric layer is 160 nm thick.

The solutions of this system are well known,[40,41] and can be expressed in terms of the refractive index of the materials. For the symmetric modes we obtain:

$$\tan\left(k_0 n_d \frac{t_d}{2}\right) = \frac{\kappa_m}{n_d} \quad ; \tag{4}$$

and for the antisymmetric ones:

$$-\cot\left(k_0 n_d \frac{t_d}{2}\right) = \frac{\kappa_m}{n_d} \qquad \qquad . \tag{5}$$

When the thickness of the metallic layer is decreased towards its skin depth, the tunneling probability of the photons through the barriers becomes significant. Therefore, the associated wavefunction accumulates a phase delay equal to the tunneling probability of the photon through one single metal barrier that can be approximated by the expression $e^{(-2k_0\kappa_m t_m)}$ (see SI section 2).[40] The analytical dispersion of such a "*leaky*" MIM can be found by adding the phase delay to the expressions for the finite square quantum well with infinite thickness of the barriers:

$$\tan\left(k_0 n_d \frac{t_d}{2} + e^{-2k_0\kappa_m t_m}\right) = \frac{\kappa_m}{n_d} \tag{6}$$

$$-\cot\left(k_0 n_d \frac{t_d}{2} + e^{-2k_0\kappa_m t_m}\right) = \frac{\kappa_m}{n_d} \tag{7}$$

Equations 6 and 7 are derived in section 2 in the SI. Figure 2c shows that the wavelength of the *quasi-bound* modes in a symmetric MIM cavity decreases with increasing metallic layer thickness until it approaches asymptotically the values of the square well system. The comparison between the quantum model, obtained by graphically solving eqs 6 and 7, and the classical treatment by Scattering Matrix Methods shows very good agreement (Fig. 2c), and validates the quantum mechanical approach.

Now we consider transmission and reflection properties of a MIM system, where the metallic layers are sufficiently thin to allow photons to tunnel through both barriers, and therefore through the entire MIM, as depicted in Figures 2b and 3a (see also section 2 in the Supporting Information). The double barrier system is schematically illustrated in Figure 3a, where the eigenmodes are depicted by solid lines, and the tunneling is sketched by the dashed lines. Tunneling through a double potential barrier with a probability approaching unity is known as *resonant tunneling*.[54–57] In real systems, however, the transmission is reduced due to losses and manifests a maximum at the resonant tunneling wavelength.[58–62] The wavelengths of the eigenmodes (from eqs 6 and 7) and the tunneling maxima as a function of the dielectric layer thickness are plotted in Figure 3b (see SI section 1 for details), together with the results for the transmittance and absorbance peaks from classical SMM calculations (as illustrated in Figure 3c), and show perfect matching. This agreement corroborates our quantum modeling, and furthermore demonstrates that the tunneling maxima correspond to the eigenmodes of the leaky quantum well, and transmittance maxima coincide with absorbance maxima, which at first seems counterintuitive. Finally, the odd modes are quenched for thin dielectric layers, when $\tan(k_0 n_d \frac{t_d}{2}) < \frac{\kappa_m}{n_d}$ (for SMM modeling and experimental spectra on the asymmetric mode see SI section 3). A very similar effect has been reported by Avrutsky et al.,[38] describing *Gap Plasmon Polaritons* (GPPs) in MIM resonators.

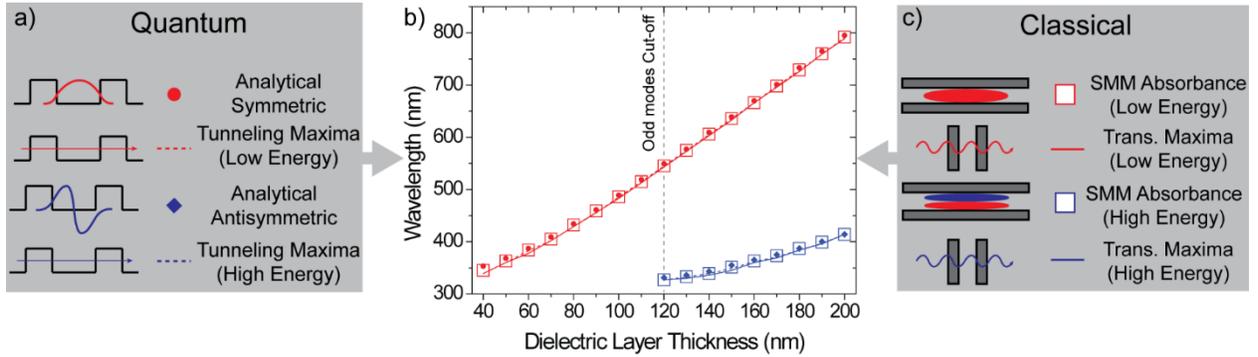

**Figure 3.** Results of the quantum modeling for quasi-bound eigenmodes and tunneling maxima (illustrated in (a)), and the transmittance and absorbance maxima obtained by classical SMM (as depicted in (c)). The respective dispersions are shown in (b) versus the thickness of the dielectric layer, demonstrating perfect agreement. The thickness of the metallic layers is 50 nm.

The MIM system can also be viewed as a single, homogenized layer corresponding to an effective potential barrier, whose height is given by its effective dielectric permittivity. A classical harmonic oscillator model can be used to derive such an effective permittivity and describe the ENZ resonances of the MIM. Due to the nanoscale thickness of the metallic layers, electrons in the metal can move freely in the plane (parallel to the interface), but are confined in the direction perpendicular to it. Just as for localized plasmons in nanoparticles, this confinement introduces a restoring force and a central oscillation frequency $\omega_{0,\text{MIM}}$ that depends on the thickness and refractive index of the dielectric material between the two metallic layers, on the thickness of the metal, and on the incidence angle. Losses, introduced by both the metal and the dielectric, result in damping denoted by $\gamma_{\text{MIM}}$. The overall optical response of the MIM can then be described as a harmonic oscillator. Under the driving action of the incident radiation, the electronic density undergoes a collective displacement $\delta$, from its equilibrium position, described by the equation of motion:

$$m_e^* \ddot{\delta}_Y + m_e^* \gamma_{\text{MIM}} \dot{\delta}_Y - m_e^* \omega_{0,\text{MIM}}^2 \delta_Y = -q\vec{E}(t);  \qquad (8)$$

where $\omega_{0,MIM}$ is the unperturbed frequency of the free electron density oscillations in the MIM, and $\delta_Y$ is the displacement of the free electron cloud in the MIM from its equilibrium position. The term $m_e^* \omega_{0,MIM}^2 \delta_Y$ is then the restoring force, and $m_e^*$ is the effective mass of the electron in the cavity.

The time dependence of the displacement $\delta_Y$ can be readily obtained:

$$\delta_Y(t) = -\frac{q\vec{E}(t)}{m_e^*(\omega^2 - \omega_{0,MIM}^2 + i\gamma_{MIM}\omega)};  \qquad (9)$$

The total dipole moment of the electrons inside the MIM cavity can then be expressed as:

$$\vec{P}_{\text{MIM}} = qN\delta_Y(t) = -\frac{Nq^2\vec{E}(t)}{m_e^*(\omega^2 - \omega_{0,MIM}^2 + i\gamma_{MIM}\omega)};  \qquad (10)$$

Where $q$ is the electronic charge, and $N$ is the density of free carriers. The electric displacement vector induced in the MIM cavity can then be seen as sustained by an effective permittivity $\varepsilon_{\text{eff,MIM}}$:

$$\vec{D}_{\text{MIM}} = \varepsilon_0\vec{E}(t) + \vec{P}_{\text{MIM}} = \varepsilon_0\varepsilon_{\text{eff,MIM}}\vec{E}(t);  \qquad (11)$$

and with eqs (10) and (11) we obtain a useful expression for $\varepsilon_{eff,MIM}$:

$$\varepsilon_{eff,MIM} = 1 - \frac{q^2 N}{m_e^* \varepsilon_0} \frac{1}{(\omega^2 - \omega_{0,MIM}^2 + i\gamma_{MIM}\omega)} ; \qquad (12)$$

$$\omega_{MIM} = \sqrt{\frac{q^2 N}{m_e^* \varepsilon_0}} \quad , \text{ and we rewrite eq (12):}$$

$$\varepsilon_{eff,MIM} = 1 - \frac{\omega_{MIM}^2}{(\omega^2 - \omega_{0,MIM}^2 + i\gamma_{MIM}\omega)} ; \qquad (13)$$

The response of the MIM is strongly related to the properties of the constituent metal, and eq 13 can then be integrated with the well-known Drude model,[63] leading to the effective permittivity of a MIM system with Ag as metal:

$$\varepsilon_{eff,MIM-Ag} = \varepsilon_\infty - \frac{\omega_p^2}{(\omega^2 + i\gamma\omega)} - \frac{\omega_{MIM}^2}{(\omega^2 - \omega_{0,MIM}^2 + i\gamma_{MIM}\omega)} ; \qquad (14)$$

here we used the classic parameters for Ag: $\gamma_{Ag}$=0.021 eV and $\omega_p$=2200 THz (corresponding to 9.1 eV).[63,64] $\omega_{0,MIM}$ and $\gamma_{MIM}$ can be measured via spectroscopic ellipsometry. Figure 4 a-c displays the experimental data obtained from a MIM cavity made of two Ag layers of 30 nm, and a central Al$_2$O$_3$ layer of 115 nm thickness. The homogenized permittivity that was measured by spectroscopic ellipsometry is compared with the effective permittivity modeled via eq 14 in Figure 4a,b. The fitting of the angles $\Psi$ and $\Delta$ in the ellipsometry measurement give $E_{0,MIM}$= 2.422 eV (~512nm), $\gamma_{MIM}$ = 0.063 eV (~ 13 nm) and $\omega_{MIM}$ =2.998 eV (~413.6nm). In order to take the residual polarizability of the MIM into account, $\varepsilon_\infty$ has to be adjusted from 5.75 (pure Ag) to 6.9. The real part of the effective dielectric permittivity in Figure 4a crosses the zero at $E_{ENZ,MIM}$ = 2.202 eV (~ 563nm). We emphasize that this ENZ resonance coincides with the symmetric mode calculated by our quantum mechanical approach via eq 6. The imaginary dielectric permittivity at this wavelength (Figure 4b) is sufficiently low to allow for an efficient coupling of the incident light with the MIM, resulting in a cavity-like resonance that behaves similar to the Ferrell-Berreman mode occurring naturally at 327 nm in Ag.[12,65] The electric field distribution for the ENZ mode, calculated by means of COMSOL-based Finite Element Method (FEM) simulations, and excited by a plane wave at incident with an angle of 40° to the normal, is shown in Figure 4d-f. The vertical (Y) component (Figure 4d) of the electric field in the cavity is responsible for the displacement $\delta_y$ of the electronic density and is at the origin of the plasmonic nature of the mode. As for the Ferrell-Berreman mode, there is no need for momentum matching techniques to excite it due to the ENZ permittivity. The in-plane (X) component shows the symmetry of the ENZ mode and relates to the waveguiding in the plane of the MIM. We note that the physical features that we obtain are consistent with the optical waveguide model given by Dionne et al.[53] Figure 4c shows that the experimental absorbance and transmission spectra have a maximum at the $ENZ_{MIM}$ wavelength, while the reflection has a minimum. This wavelength corresponds also to the analytically calculated symmetric mode and to the tunneling maximum of the MIM. We therefore conclude that the cavity modes occurring under p-polarised light with oblique angle of incidence in a subwavelength MIM are Ferrell-Berreman resonances that correspond to ENZ resonant tunneling modes. The resonance frequency of these modes can be engineered by tuning the thickness of the layers composing the MIM, and through the choice of the constituent materials.

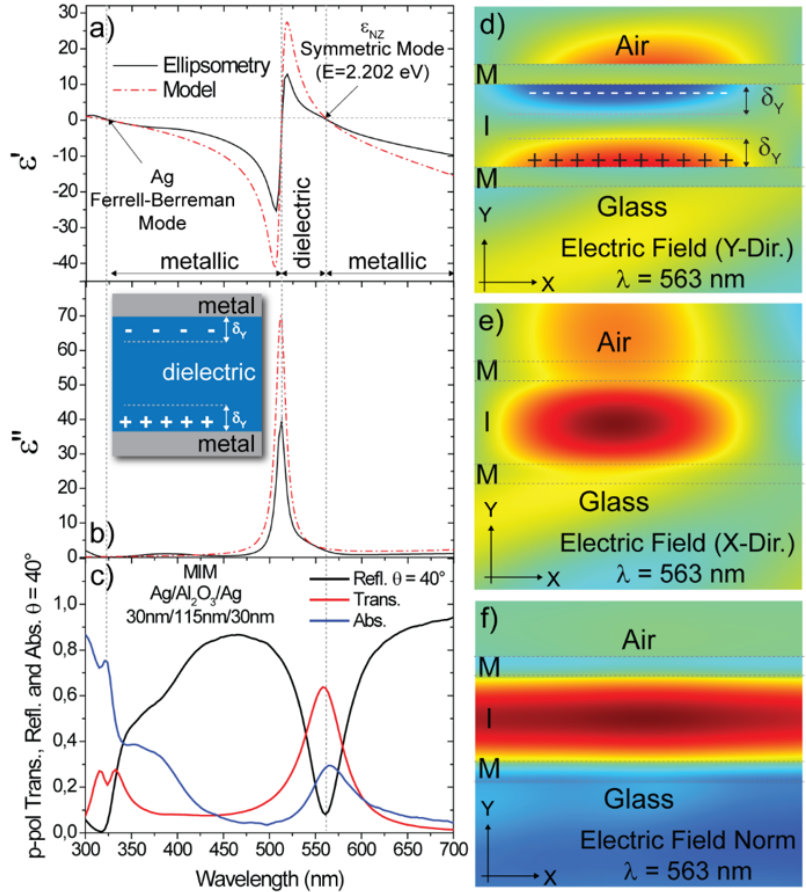

**Figure 4.** Experimental ellipsometrically measured (black solid curves) and theoretically modeled (red dashed curves) real (a) and imaginary (b) dielectric permittivity, detected at 40°, of a symmetric MIM structure made of Ag = 30 nm and Al₂Oᵢ = 115 nm. The curves show perfect agreement on the Ag Ferrell-Berreman mode and the ENZ symmetric mode at E=2.202 eV. (c) Experimentally measured absorbance (blue curve), reflectance (black curve) and transmittance (red curve). Inset in (b) illustrates the MIM structure, and indicates the charge displacement in the dielectric layer. COMSOL simulation of the vertical (d) and horizontal (e) distribution of the electric field in the MIM at the resonant wavelength, calculated at 40°, together with the simulated norm of the electric field (f) calculated as $\sqrt{E_X^2 + E_Y^2}$.

With the approach of an effective permittivity, the MIM structure can be treated as an artificial, homogenized layer that can have either metallic ($\varepsilon_{\text{eff,MIM}} < 0$) or dielectric ($\varepsilon_{\text{eff,MIM}} > 0$) properties, or which acts as an ENZ layer when $\varepsilon_{\text{eff,MIM}} = 0$. This behavior, seen as photons impinging on the effective MIM has insightful quantum mechanical analogies. By comparing eqs 1&2, we see that the potential energy U(x) is the equivalent of $-(n_j - \kappa_j)^2$. By fixing $U(x) = V_0$ we can express the optical potential in terms of wavevector as $k^2/k_0^2$, and three scenarios can be compared, as illustrated in Figure 5.

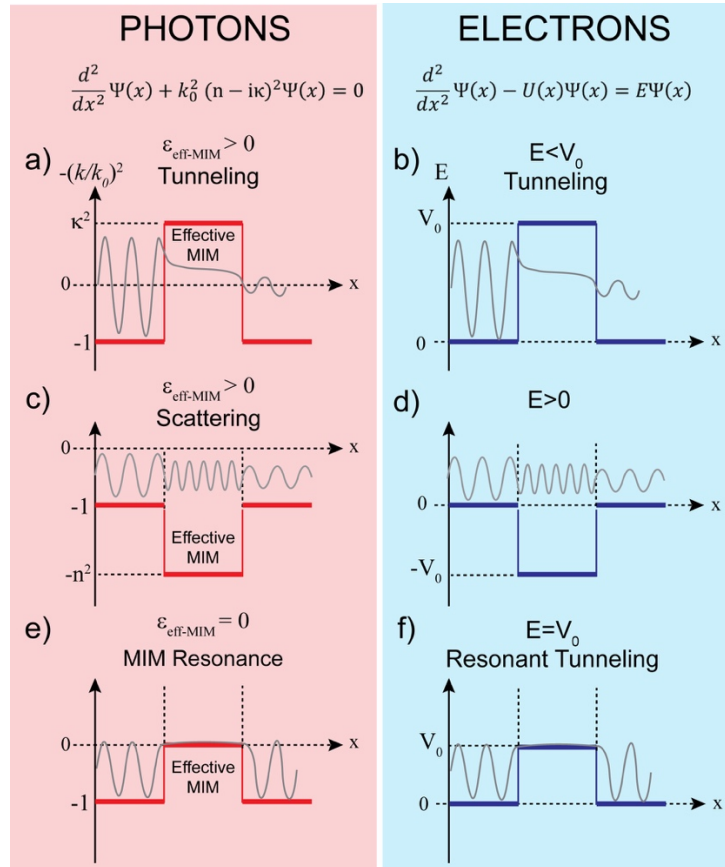

**Figure 5.** Quantum mechanical analogies of the MIM structure, treated as a homogenized layer. Photon tunneling through a metal (a) corresponds to electron tunneling through a potential barrier (b). Photons scattered on a dielectric medium (c) correspond to electrons passing over a quantum well (d), and photons impinging on an ENZ layer at resonance (e) can be seen as the equivalent of resonant tunneling of electrons through a potential barrier (f). Wavefunctions are sketched in grey with arbitrary scale of amplitude.

The photonic case of a MIM as an effective metal corresponds to electron tunneling through a potential barrier (Fig. 5a,b). Here $k$ is purely imaginary in both cases and the wavefunctions are evanescent waves. The MIM as an effective dielectric is analogous to electrons scattered on a potential well (Fig. 5c,d), the wavevector is positive and the wavefunctions are real propagating waves. The case of $\varepsilon'_{eff-MIM} \approx 0$ corresponds to electrons with an energy equal to the barrier height (Fig. 5e,f), consequently the wavefunction is constant inside the barrier, which enables resonant tunneling.

In this case, it is straightforward to conclude that the wavevector for an electron is zero, since $E=V_o$. In analogy, for photons propagating inside the effective MIM the wavefunction is equal to $\Psi_{eff}(x) = A\, e^{ik_{eff}x}$. Therefore:

$$k_{eff} = k_0 n_{eff\_MIM} = k_0 \sqrt{\varepsilon'_{eff\_MIM} - i\varepsilon''_{eff\_MIM}} = 0 \tag{15}$$

And for a low imaginary part (which is always the case for the considered resonances):

$$k_0 \sqrt{\varepsilon'_{eff\_MIM}} = 0 \rightarrow \varepsilon'_{eff\_MIM} = 0 \tag{16}$$

We demonstrated in Figure 3 that the resonant tunneling wavelengths coincide with the quasi-bound modes of the leaky cavity. Moreover, in correspondence of these modes a near-zero permittivity has been experimentally measured. Equation 16 reveals that a vanishing permittivity naturally leads to a zero wavevector that reduces the wavefunction to a constant within the effective barrier constituted by the homogenized MIM. We can therefore conclude that the resonances of the MIM correspond to resonant tunneling modes in which the wavevector is zero and the wavefunction of the photon can be seen to propagate as a constant through the homogenized MIM. This finding also elucidates why transmittance and absorbance maxima in MIM systems occur at the same wavelength, since both effects are associated to the cavity resonances. We note that the cavity modes can be excited both with p- and s-polarized light, however, mode frequency and line shape are slightly different for the two polarizations, as discussed in detail in section 3 of the SI. Furthermore, only resonances excited under oblique angles and with p-polarized light can be associated to Ferrell-Berreman modes, while ENZ cavity resonances occur also under normal incidence and with s-polarized light.

A special case of an asymmetric MIM structure with a very thin top, and very thick bottom metal layer acts as a superabsorber. In the following, we will show that the quantum treatment is also capable of describing this, from an application point highly relevant, structure. Figure 6a shows the experimental reflectance of such a structure, where the top Ag layer has a thickness of 40 nm, the dielectric of 160 nm, and the bottom Ag layer acting as backreflector is 200 nm thick. Two pronounced minima in reflectance can be identified. We note that the minimum at small wavelengths corresponds to the odd mode that appears only for cavities with a thickness of the dielectric layer that exceeds a threshold value. We will show in the following that these minima in reflectance correspond to the ENZ modes of the MIM cavity, and to the quasi-bound eigenmodes of the non-symmetric finite potential quantum well. Interestingly, a single Ag layer with the same thickness as the top layer of the superabsorber MIM (40 nm) is highly reflective throughout the visible range, as depicted in Figure 6a by the blue dashed line. The appearance of a reflectance minimum, like the one at 670 nm, in the presence of successive layers points again to the quantum analogy that resembles tunneling behavior.

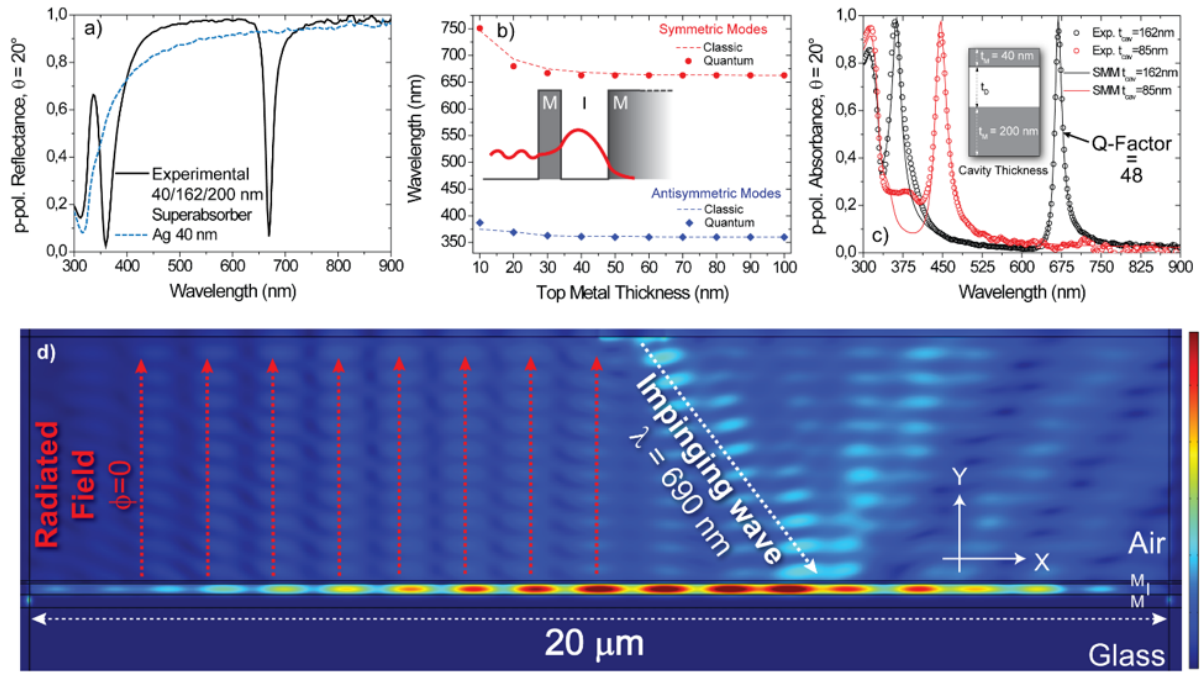

**Figure 6.** (a) Experimental p-polarized reflectance recorded from a 40 nm Ag film (blue dashed curve) and from a 40/162/200 nm Ag/Al$_2$O$_3$/Ag MIM. (b) Even (red circles) and odd modes (blue diamonds), calculated as a function of the top Ag layer thickness via eqs 14 and 15, compared with the SMM calculated even (red dashed curve) and odd (blue dashed curve) modes. (Al$_2$O$_3$ = 150 nm). The inset sketches the shape of the even mode profile. (c) Measured (dots) and SMM simulated (solid lines) absorbance for two superabsorbers (t$_{Al2O3}$=162 nm (black curve) and t$_{Al2O3}$=85 nm (red curve)) where the absorbance is obtained as 1-transmittance-reflectance. (d) Electric field norm calculated via FEM simulations (COMSOL Multiphysics), for a 40nm/170nm/200nm Ag/Al$_2$O$_3$/Ag MIM superabsorber, θ=40° (white dashed line arrow), at the cavity mode (690 nm). The red arrows are a guide to the eye, indicating the radiation direction perpendicular to the surface plane. Red (blue) color corresponds to high (low) electric field.

Although the symmetry in the superabsorber MIM structure is broken, we can find equations for the quasi-bound eigenmodes and apply our quantum mechanical model as follows: by considering an infinitely thick barrier for the backreflector in our model, as illustrated in Figure 6b, we can allocate the entire phase shift introduced by the tunneling to the thin barrier. Since the photon tunnels twice (in and out), the phase shift is twice as large as in the one in eqs 6 and 7 that were derived for symmetric barriers. With this consideration, we obtain the analytic dispersions for the symmetric and anti-symmetric modes as:

$$\tan\left(k_0 n_j \frac{t_d}{2} + e^{-4k_0 \kappa_m t_m}\right) = \frac{\kappa_m}{n_d}; \tag{17}$$

$$-\cot\left(k_0 n_j \frac{t_d}{2} e^{-4k_0 \kappa_m t_m}\right) = \frac{\kappa_m}{n_d} \tag{18}$$

Figure 6b shows very good agreement between the analytical (quantum mechanical, eqs 17 and 18) and the numerical (SMM-classical) calculated eigen-(cavity) modes as a function of the thickness of the top metal layer of a superabsorber MIM with a 150 nm dielectric layer of Al$_2$O$_3$. The resonance wavelength of a superabsorber can be tuned via the thickness of the dielectric layer. Figure 6c plots the ellipsometrically measured and SMM simulated absorbance for dielectric layers with 85 nm and 162 nm thickness. The experimentally detected absorbance is above 95% for all the resonances, even and odd. The Q-factor of the even mode at long wavelength is 48, which is a very high value for plasmonic and ENZ resonances.[66-68] Figure 6d shows the electric field distribution at resonance under excitation with a monochromatic wave at λ = 690 nm incident at an angle of 40° as depicted by the white dashed arrow. Interestingly, we observe a lateral (X-direction) distribution of the electric field for several microns, which indicates that the photons at the ENZ mode propagate in the cavity, which might be related to slow light trapping in ENZ materials, as reported by Ciattoni et al.[65] Following their analysis, we take the dispersion relation for transverse plane waves, $k(\omega) = (\omega/c \sqrt{\varepsilon(\omega)}$, and calculate the group velocity $v_g(\omega) =$ $d\omega/dk$. With the relation for the dielectric permittivity of the effective MIM at resonance in eq 14, we can express the group velocity as:

$$v_g(\omega) = c \left[ \sqrt{\varepsilon_{eff-MIM}(\omega)} + \frac{\omega}{\sqrt{\varepsilon_{eff-MIM}(\omega)}} \alpha(\omega) \right]^{-1}, \qquad (19)$$

where $\alpha(\omega) = \frac{\omega_p^2}{\omega^3} - \frac{\omega \omega_{MIM}^2}{(\omega^2 - \omega_{0,MIM}^2)^2}$ . Eq 19 shows that the group velocity goes to zero when $\varepsilon_{eff-MIM}(\omega) \sim 0$, which reveals slow-light propagation in the lateral directions.

The phase contribution given by the incident wave at oblique angle is nullified by the cavity resonance, and generates a radiation field outside the cavity propagating perpendicular to the surface plane. This phenomenon is more evident on the left side in Figure 6d, because the radiation there does not interfere with the Snell reflected beam from the surface, and such propagation of the radiated field with a phase front parallel to the interface is a characteristic feature of ENZ media that can be used for re-shaping of the phase-front.[2]

The tunnel effect in a MIM *superabsorber* is very sensitive to the refractive index of the material on the top thin (*tunneling*) metal. Therefore, the MIM *superabsorber* can be used as a refractive index sensor, by considering a thin dielectric layer on the top of it that causes an additional phase delay. An exact expression for the correction factor to describe this additional phase delay can be obtained by solving the system of equations that account for reflection and transmission at the barriers in the system, as illustrated in Figure 7a and detailed in section 4 of the SI. However, SMM simulations show that the phase delay caused by the sensing layer is roughly proportional to its thickness and refractive index, and inversely proportional to the thickness of the tunneling metal. Therefore, we approximate the phase change by the correction factor $\varphi_s = k_0 n_s\, t_s f_s$ , where $t_s$ is the thickness of the sensing layer, $n_s$ its refractive index and $f_s$ its fill fraction expressed as $t_s/(t_s+t_m)$ with $t_m$ being the thickness of the top metal layer (see Supporting Information). This additional phase delay introduced by the sensing layer results in a red shift of the ENZ wavelength of the MIM, and consequently to a red shift of the sharp absorbance peak. The MIM superabsorber-based refractive index sensor has two fundamental advantages: (i) the operating spectral band can be tuned by the thickness of core dielectric layer of the MIM throughout the visible range, and (ii) relies on a very simple fabrication that can be readily integrated on the apex of optical fibers. Figure 7b shows the refractive index dependence of the wavelength of the ground (even) mode absorbance peak for sensing layers with thickness of 10 nm, 30 nm, 50 nm, where the MIM consists of a 10 nm thick Ag top layer, 150 nm of $Al_2O_3$ as dielectric, and a Ag backreflector with 200 nm thickness. The sensitivity of plasmonic sensors is typically evaluated as the spectral shift of the plasmonic peak per *refractive index unit* (RIU): *S=Δλ/Δn_s [nm/RIU]*.[69–73] The response in Figure 7b is slightly non-linear, with a higher sensitivity for higher refractive index, which results in different working regions: for $1<n_s<2.2$, typical of the most common oxides, the sensitivity is around 25 nm/RIU, for refractive index higher than 3 it is around 45 nm/RIU. The sensitivity depends also on the top metal layer thickness, and can be tuned to more linear behavior with a slightly reduced sensitivity with an increased metal thickness, of for example 20 nm. We note that the performance of the superabsorber refractive index sensor cannot compete with highly sophisticated metamaterial sensors tailored for biomolecule detection as reported in literature.[74,75] However, our system can find an application range as sensors in thin film technologies, where larger changes in refractive index should be detected, or the effective refractive index of a composite material should be measured, for example of a nanocrystal solid, or where the layer thickness of a dielectric with a known refractive index is of interest.

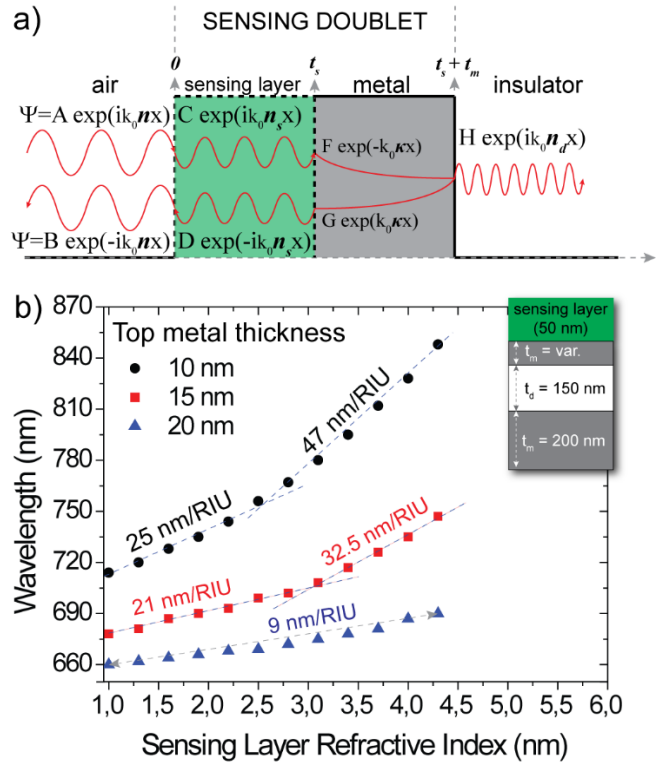

**Figure 7**. (a) Sketch of the reflection, transmission and tunneling process at the top metal layer of a superabsorber coated with a thin dielectric (sensing) layer. (b) Wavelength dependence of the even mode resonance on the refractive index of the sensing layer for a MIM with 10 nm Ag top layer, 150 nm Al$_2$O$_3$ core, and a 200 nm thick Ag backreflector. Dispersions for three different thickness of the sensing layer that are typical in thin film technology are shown.

Conclusions

We analyzed the MIM system with a quantum mechanical approach that reveals the cavity modes as even and odd eigenmodes of a quantum well system. This treatment enables analytical solutions for the resonance frequencies in two practically highly relevant cases: symmetric MIMs, and MIMs with one very thick metal layer that allows no transmission. Therefore, the analytical treatment provides an intuitive alternative pathway to numerical simulations commonly used in electromagnetic theory. Furthermore, the quantum view provides new insight in the physics of MIM cavities. We empirically defined a hermiticity limit that reveals why ENZ resonances are forbidden in spectral ranges where the interband transitions of the metals occur. In contrast to surface plasmon polaritons at metal/insulator interfaces, the cavity resonances in the MIM are ENZ modes that can be excited without any need of momentum matching. In the case of p-polarized light under oblique incidence angle the resonances have a plasmonic nature and can be associated to Ferrell-Berreman modes. The models we presented, and the analogy of the MIM as the optical counterpart to the electronic tunnel diode, will be highly versatile for the design and understanding of complex metal-insulator structures such as multiple stacked MIM cavities. On one hand, this provides a design toolbox, while on the other hand it reveals underlying physical processes that can be exploited in novel devices. Our work elucidated how the dispersion of the permittivity can be tuned in MIM structures, therefore building the basis for the design of artificial metals featuring engineered ENZ bandgaps.

Materials and Methods

*Fabrication and characterization of the MIM structures.*

The MIM structures have been fabricated by a multistep process that consists of deposition of (i) the metal (Ag), and (ii) the dielectric ($Al_2O_3$) layers.  For (i), electron-beam induced thermal evaporation (Kurt J. Lesker PVD 75) of Ag on a glass substrate was employed to obtain a Ag layer with desired thickness, then followed by deposition of 10 nm $Al_2O_3$ inside the same system to prevent Ag from oxidation.  For (ii), the $Al_2O_3$ was deposited in an atomic layer deposition (ALD) system (FlexAl from Oxford Instruments) using a thermal deposition process with a stage temperature of 110 °C, resulting in an alumina deposition rate of 0.09 nm/cycle. Tri-methylaluminate (TMA) and $H_2O$ were used as precursors. A heating step of 300 s was performed before starting the ALD cycles. Each ALD cycle consisted of a $H_2O$/purge/TMA/purge sequence with a pulse durations of 0.075/6/0.033/2 seconds, respectively.

The characterization of the optical properties of all the fabricated multilayer structures has been performed by spectroscopic Ellipsometry with a Vertical Vase ellipsometer by Woollam, in the range from 300-900 nm. Spectroscopic analysis has recorded at three different angles (50°, 60° and 70°) with a step of 3 nm. P-Polarized reflectance and transmittance measurements have been measured via ellipsometry as well, in a broad range of angles, in which falls the case of 40° discussed in this work. The resolution of recorded spectra is 3 nm, and all spectra have been normalized to the intensity of the Xe lamp.

***Modeling and Simulations.***

Finite Element Method based full-field simulations have been carried out by means of COMSOL Multiphysics. As boundary conditions, appropriately swept-meshed Perfectly Matched Layers have been imposed, while the nanometric layers constituting the MIMIM have been meshed with a free triangular texture in which the maximum element size is 2 nm. Illumination is provided by a plane wave impinging at 40° via a classic internal port excitation (port width is 2 $\mu$m), adjacent to the upper Perfectly Matched Layer domain. The dielectric permittivity of the selected materials was taken from experimentally measured data. Scattering Matrix Method simulations have been conducted by means of a customized MATLAB code.

ASSOCIATED CONTENT

**Supporting Information**. Tunneling of photons through a potential barrier; Bound modes and quasi bound modes of the MIM system; Resonant tunneling through a MIM cavity and ENZ nature of the resonant tunneling modes; Analytical dispersion relation of the resonant tunneling refractive index sensor. The following files are available free of charge. MIM_SI.pdf

Author Contributions. The manuscript was written through contributions of all authors. All authors have given approval to the final version of the manuscript.

ACKNOWLEDGMENT

We thank Francesco Plastina and Andrea Marini for fruitful discussions. The research leading to these results has received funding from the European Union under the Marie Skłodowska-Curie Grant Agreement COMPASS No. 691185.

REFERENCES

(1)     Engheta, N. *Science* **2013**, *340*, 286–288.

(2)     Alù, A.; Silveirinha, M. G.; Salandrino, A.; Engheta, N. *Phys. Rev. B - Condens. Matter Mater. Phys.* **2007**, *75*, 155410.

(3)     Monticone, F.; Doeleman, H. M.; Den Hollander, W.; Koenderink, A. F.; Alù, A. *Laser Photonics Rev.* **2018**, *12*, 1700220.

(4)     Khademi, A.; Dewolf, T.; Gordon, R. *Opt. Express* **2018**, *26*, 15656–15664.

(5)     Passler, N. C.; Gubbin, C. R.; Folland, T. G.; Razdolski, I.; Katzer, D. S.; Storm, D. F.; Wolf, M.; De Liberato, S.; Caldwell, J. D.; Paarmann, A. *Nano Lett.* **2018**, *18*, 4285–4292.

(6)     Halterman, K.; Alidoust, M.; Zyuzin, A. *Phys. Rev. B* **2018**, *98*, 085109.

(7)     Krasikov, S.; Iorsh, I. V.; Shalin, A.; Belov, P. A. *Phys. Status Solidi RRL* **2014**, *8*, 1015–1018.

(8)     Savoia, S.; Castaldi, G.; Galdi, V.; Alù, A.; Engheta, N. *Phys. Rev. B - Condens. Matter Mater. Phys.* **2015**, *91,115114.*

(9)     Silveirinha, M. G.; Engheta, N. *Phys. Rev. B - Condens. Matter Mater. Phys.* **2007**, *76*, 245109.

(10)   Feng, S.; Halterman, K. *Phys. Rev. B - Condens. Matter Mater. Phys.* **2012**, *86*, 165103.

(11)   Anopchenko, A.; Tao, L.; Arndt, C.; Lee, H. W. H. *ACS Photonics* **2018**, *5*, 2631–2637.

(12)   Newman, W.; Cortes, C. L.; Atkinson, J.; Pramanik, S.; DeCorby, R. G.; Jacob, Z. *ACS Photonics* **2014**, *2*, 2–7.

(13)   Ferrell, R. A.; Stern, E. A. *Am. J. Phys.* **1962**, *30*, 810-812.

(14)   Berreman, D. W. *Phys. Rev.* **1963**, *130*, 2193.

(15)   Ferrell, R. A. *Phys. Rev.* **1958**, *111*, 1214.

(16)   Pollard, R. J.; Murphy, A.; Hendren, W. R.; Evans, P. R.; Atkinson, R.; Wurtz, G. A.; Zayats, A. V., Podolskiy, A. V. *Phys. Rev. Lett.* **2009**, *102*, 127405.

(17)   Caspani, L.; Kaipurath, R. P. M. P. M.; Clerici, M.; Ferrera, M.; Roger, T.; Kim, J.; Kinsey, N.; Pietrzyk, M.; Di Falco, A.; Shalaev, V. M. M.; Boltasseva, A.; Faccio, D. *Phys. Rev. Lett.* **2016**, *116*, 233901.

(18)    Argyropoulos, C.; Chen, P.-Y.; D'Aguanno, G.; Engheta, N.; Alù, A. *Phys. Rev. B* **2012**, *85*, 045129.

(19)    Alam, M. Z.; De Leon, I.; Boyd, R. W. *Science* **2016**, *352*, 795–797.

(20)    Hendrickson, J. R.; Vangala, S.; Dass, C.; Gibson, R.; Goldsmith, J.; Leedy, K.; Walker, D. E.; Cleary, J. W.; Kim, W.; Guo, J. *ACS Photonics* **2018**, *5*, 776–781.

(21)    Mrabti, A.; Lévêque, G.; Akjouj, A.; Pennec, Y.; Djafari-Rouhani, B.; Nicolas, R.; Maurer, T.; Adam, P.-M. *Phys. Rev. B* **2016**, *94*, 075405.

(22)    Kim, J.; Carnemolla, E. G.; DeVault, C.; Shaltout, A. M.; Faccio, D.; Shalaev, V. M.; Kildishev, A. V.; Ferrera, M.; Boltasseva, A. *Nano Lett.* **2018**, *18*, 740–746.

(23)    Schaffernak, G.; Krug, M. K.; Belitsch, M.; Gasparic, M.; Ditlbacher, H.; Hohenester, U.; Krenn, J. R.; Hohenau, A. *ACS Photonics* **2018**, *5*, 4823-4827.

(24)    Hackett, L. P.; Ameen, A.; Li, W.; Dar, F. K.; Goddard, L. L.; Liu, G. L. *ACS Sens.* **2018**, *3*, 290–298.

(25)    Verre, R.; Yang, Z. J.; Shegai, T.; Käll, M. *Nano Lett.* **2015**, *15*, 1952–1958.

(26)    Tagliabue, G.; Poulikakos, D.; Eghlidi, H. *Appl. Phys. Lett.* **2016**, *108*, 221108.

(27)    Limonov, M. F.; Rybin, M. V.; Poddubny, A. N.; Kivshar, Y. S. *Nat. Photonics* **2017**, *11*, 543–554.

(28)    Meng, Z. M.; Qin, F. *Plasmonics* **2018**, *13, 2329-2336*.

(29)    Shaltout, A. M.; Kim, J.; Boltasseva, A.; Shalaev, V. M.; Kildishev, A. V. *Nat. Commun.*

**2018**, *9*, 2673–2680.

(30)     Dionne, J. A.; Verhagen, E.; Polman, A.; Atwater, H. A. *Opt. Express* **2008**, *16*, 19001–19017.

(31)     Lezec, H. J.; Dionne, J. A.; Atwater, H. A. *Science* **2007**, *316*, 430–432.

(32)     Dionne, J. A.; Sweatlock, L. A.; Atwater, H. A.; Polman, A. *Phys. Rev. B - Condens. Matter Mater. Phys.* **2005**, *72*, 075405.

(33)     Kurokawa, Y.; Miyazaki, H. T. *Phys. Rev. B - Condens. Matter Mater. Phys.* **2007**, *75*, 035411.

(34)     Li, Z.; Butun, S.; Aydin, K. *ACS Photonics* **2015**, *2*, 183–188.

(35)     Caligiuri, V.; Palei, M.; Imran, M.; Manna, L.; Krahne, R. *ACS Photonics* **2018**, *5*, 2287–2294.

(36)     Li, L.; Wang, W.; Luk, T. S.; Yang, X.; Gao, J. *ACS Photonics* **2017**, *4*, 501–508.

(37)     Maier, S. A. *Plasmonics: Fundamentals and Applications*; Springer, New York **2007**.

(38)     Avrutsky, I.; Salakhutdinov, I.; Elser, J.; Podolskiy, V. *Phys. Rev. B - Condens. Matter Mater. Phys.* **2007**, *75*, 241402.

(39)     R. Eisberg and R. Resnick. *Quantum Physics of Atoms, Molecules, Solids, Nuclei, and Particles*; John Wiley and Sons, New York **1985**.

(40)     Griffiths, D. J. *Introduction to quantum mechanics*, Cambridge University Press, Cambridge (UK) **2005**.

(41)   Schwabl, F. *Quantum Mecanics*; Springer, Berlin **2007**.

(42)   Rüter, C. E.; Makris, K. G.; El-Ganainy, R.; Christodoulides, D. N.; Segev, M.; Kip, D. *Nat. Phys.* **2010**, *6*, 192–195.

(43)   Makris, K. G.; El-Ganainy, R.; Christodoulides, D. N.; Musslimani, Z. H. *Phys. Rev. Lett.* **2008**, *100*, 103904.

(44)   Marte, M. A. M.; Stenholm S. *Phys. Rev. A* **1997**, *56*, 2940.

(45)   Alaeian, H.; Dionne, J. A. *Phys. Rev. B - Condens. Matter Mater. Phys.* **2014**, *89*, 075136.

(46)   Feng, L.; El-Ganainy, R.; Ge, L. *Nat. Photonics* **2017**, *11*, 752–762.

(47)   Hodaei, H.; Miri, M.; Heinrich, M.; Christodoulides, D. N.; Khajavikhan, M. *Science.* **2014**, *346*, 975-978.

(48)   Johnson, P. B.; Christy, R. W. *Phys. Rev. B* **1972**, *6*, 4370.

(49)   Born, M.; Wolf, E. *Principles of Optics Electromagnetic Theory of Propagation, Interference and Diffraction of Light,* Pergamon Press, Oxford (UK) **1980**.

(50)   Caligiuri, V.; Pezzi, L.; Veltri, A.; De Luca, A. *ACS Nano* **2017**, *11*, 1012–1025.

(51)   Caligiuri, V.; De Luca, A. *J. Phys. D Appl. Phys.* **2016**, *49*, 08LT01.

(52)   Sorger, V. J.; Oulton, R. F.; Yao, J.; Bartal, G.; Zhang, X. *Nano Lett.* **2009**, *9*, 3489–3493.

(53)   Dionne, J. A.; Sweatlock, L. A.; Atwater, H. A.; Polman, A. *Phys. Rev. B - Condens. Matter Mater. Phys.* **2006**, *73*, 035407.

(54)   Weil, T.; Vinter, B. *Appl. Phys. Lett.* **1987**, *50*, 1281–1283.

(55) Ricco, B.; Azbel, M. Y. *Phys. Rev. B* **1984**, *29*, 1970-1981.

(56) Sollner, T. C. L. G.; Goodhue, W. D.; Tannenwald, P. E.; Parker, C. D.; Peck, D. D. *Appl. Phys. Lett.* **1983**, *43*, 588–590.

(57) Chang, L. L.; Esaki, L.; Tsu, R. *Appl. Phys. Lett.* **1974**, *24*, 593–595.

(58) Castaldi, G.; Galdi, V.; Alù, A.; Engheta, N. *J. Opt. Soc. Am. B* **2011**, *28*, 2362–2368.

(59) Moccia, M.; Castaldi, G.; Galdi, V.; Alù, A.; Engheta, N. *J. Phys. D. Appl. Phys.* **2013**, *47*, 025002.

(60) Moccia, M.; Castaldi, G.; Galdi, V.; Alù, A.; Engheta, N. *J. Appl. Phys.* **2014**, *115*, 043107.

(61) Castaldi, G.; Gallina, I.; Galdi, V.; Alù, A.; Engheta, N. *Phys. Rev. B - Condens. Matter Mater. Phys.* **2011**, *83*, 081105(R).

(62) Zhou, L.; Wen, W.; Chan, C. T.; Sheng, P. *Phys. Rev. Lett.* **2005**, *94*, 243905.

(63) Kittel, C. *Introduction to solid state physics*, John Wiley and Sons, New York **1996**.

(64) Cai, W.; Shalaev, V. *Optical metamaterials*; Springer-Verlag New York, **2010**.

(65) Ciattoni, A.; Marini, A.; Rizza, C.; Scalora, M.; Biancalana, F. *Phys. Rev. A - At. Mol. Opt. Phys.* **2013**, *87*, 053853.

(66) Savasta, S.; Saija, R.; Ridolfo, A.; Di Stefano, O.; Denti, P.; Borghese, F. *ACS Nano* **2010**, *4*, 6369–6376.

(67) MacDonald, K. F.; Sámson, Z. L.; Stockman, M. I.; Zheludev, N. I. *Nat. Photonics* **2009**, *3*, 55–58.

(68)  Lehmann, J.; Merschdorf, M.; Pfeiffer, W.; Thon, A.; Voll, S.; Gerber, G. *Phys. Rev. Lett.* **2000**, *85*, 2921–2924.

(69)  Mazzotta, F.; Höök, F.; Jonsson, M. P. *Nanotechnology* **2012**, *23*, 415304.

(70)  Martín-Becerra, D.; Armelles, G.; González, M. U.; García-Martín, A. *New J. Phys.* **2013**, *15*, 085021.

(71)  Shen, H.; Lu, G.; Zhang, T.; Liu, J.; Gu, Y.; Perriat, P.; Martini, M.; Tillement, O.; Gong, Q. *Nanotechnology* **2013**, *24*, 285502.

(72)  Sherif, S. M.; Zografopoulos, D. C.; Shahada, L. A.; Beccherelli, R.; Swillam, M. *J. Phys. D. Appl. Phys.* **2017**, *50*, 055104.

(73)  Wadell, C.; Syrenova, S.; Langhammer, C. *ACS Nano* **2014**, *8*, 11925–11940.

(74)  Kabashin, A. V.; Evans, P.; Pastkovsky, S.; Hendren, W.; Wurtz, G. A.; Atkinson, R.; Pollard, R.; Podolskiy, V. A.; Zayats, A. V. *Nat. Mater*. **2009**, *8*, 867–871.

(75)  Sreekanth, K. V.; Alapan, Y.; ElKabbash, M.; Ilker, E.; Hinczewski, M.; Gurkan, U. A.; De Luca, A.; Strangi, G. *Nat. Mater*. **2016**, *15*, 621–627.

SYNOPSIS (Word Style "SN_Synopsis_TOC").

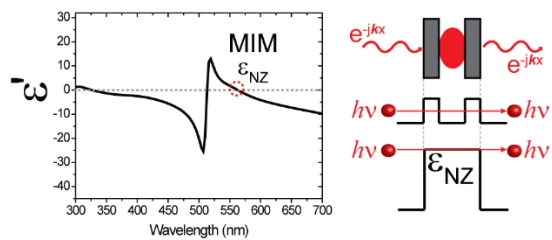